\documentclass[twocolumn,prl,superscriptaddress,showpacs,floatfix]{revtex4}
\usepackage{graphicx}
\usepackage{amsmath}
\usepackage{amssymb}

\newcommand{\species}[1]{{\it #1}}
\newcommand{\operon}[1]{{#1}}

\newcommand{\TF}[1]{{\mathrm{#1}}}
\newcommand{\TFA}{\TF{A}}
\newcommand{\TFB}{\TF{B}}

\newcommand{\NA}{N_{\TFA}}
\newcommand{\NB}{N_{\TFB}}

\newcommand{\OO}{{\mathrm{O}}}

\newcommand{\nothing}{\emptyset}
\newcommand{\nmean}{\overline N}

\newcommand{\kf}{k_{\mathrm{f}}}
\newcommand{\kb}{k_{\mathrm{b}}}
\newcommand{\kon}{k_{\mathrm{on}}}
\newcommand{\koff}{k_{\mathrm{off}}}
\newcommand{\kA}{k_{\TFA}}
\newcommand{\kB}{k_{\TFB}}
\newcommand{\muA}{\mu_{\TFA}}
\newcommand{\muB}{\mu_{\TFB}}
\newcommand{\Kd}{K_{\mathrm{d}}}
\newcommand{\Kb}{K_{\mathrm{b}}}
\newcommand{\lphage}{$\lambda$-phage}

\newcommand{\tick}{\checkmark}
\newcommand{\cross}{$\times$}

\begin{document}

\title{Enhancement of the stability of genetic switches by\\ 
overlapping upstream regulatory domains}

\author{Patrick B. Warren}
\affiliation{FOM Institute for Atomic and Molecular Physics,
Kruislaan 407, 1098 SJ Amsterdam, The Netherlands.}
\affiliation{Unilever R\&D Port Sunlight,
Bebington, Wirral, CH63 3JW, UK.}

\author{Pieter Rein ten Wolde}
\affiliation{FOM Institute for Atomic and Molecular Physics,
Kruislaan 407, 1098 SJ Amsterdam, The Netherlands.}
\affiliation{Division of Physics and Astronomy, Vrije Universiteit, De
Boelelaan 1081, 1081 HV Amsterdam, The Netherlands.}
\date{October 31, 2003}

\begin{abstract}
We study genetic switches formed from pairs of mutually repressing
operons.  The switch stability is characterised by a well defined
lifetime which grows sub-exponentially with the number of
copies of the most-expressed transcription factor, in the regime
accessible by our numerical simulations.  The stability can be
markedly enhanced by a suitable choice of overlap between the upstream
regulatory domains.  Our results suggest that robustness against
biochemical noise can provide a selection pressure that drives
operons, that regulate each other, together in the course of
evolution.
\end{abstract}

\pacs{87.16.Yc; 05.40.-a}


\maketitle 

Biochemical networks are the analog computers of life. They allow
living cells, to detect, transmit, and amplify environmental signals,
as well as integrate different signals in order to recognize patterns
in, say, the food supply. Indeed, biochemical networks can perform a
variety of computational tasks analogous to electronic
circuits. However, their design principles are markedly different. In
a biochemical network, computations are performed by molecules that
chemically and physically interact with each other. These interactions
are stochastic in nature. This becomes particularly important when the
concentrations are low. In gene regulatory networks, this is generally
the case: not only the DNA, but also the proteins that regulate gene
expression are often present in very small numbers, which can be as
low as ten, or even fewer. Hence, one would expect that gene
regulatory networks, in contrast to electronic circuits, are highly
stochastic and error
prone~\cite{McAA,Elowitz,Barkai,Ozbudak,Elowitz02}. An important
question, therefore, is how the ability to resist biochemical noise
constrains the design of the network~\cite{Elowitz,Barkai}.

In prokaryotes, the expression of operons---groups of contiguous genes
that are transcribed into single mRNA molecules---is regulated by the
binding of transcription factors (TFs) to upstream regulatory domains
on the DNA.  A spatial arrangement in which two operons are
transcribed in diverging directions allows the upstream regulatory
domains to interfere with each other. This affords additional
regulatory control. In particular, biochemical noise in the expression
of operons can become correlated or anticorrelated. Just as the
existence of operons provides for correlated \emph{gene} expression,
interference between the regulatory domains of two diverging operons
allows for correlated or \emph{anticorrelated} expression of
\emph{operons}. Here, we show that this can have a dramatic influence
on the stability of gene regulatory networks.

\begin{figure}
\begin{center}
\includegraphics{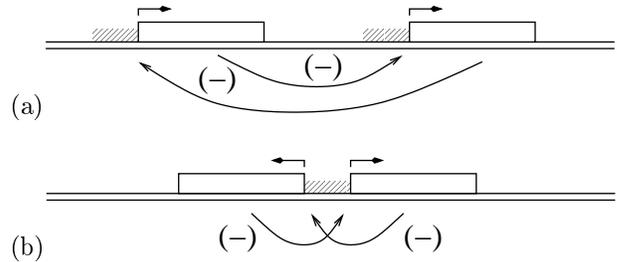}
\end{center}
\caption{(a) In a prokaryote, a so-called toggle switch can be formed
from a pair of operons that mutually repress each other. (b) If these
operons are transcribed in diverging orientations, the upstream
regulatory domains can interfere with each other. \label{switchfig}}
\end{figure}

Recently, we have performed a statistical analysis of the spatial
distribution of operons on the genome of \species{Escherichia
coli}~\cite{colistat}. The analysis identified a large number of
motifs in which the operator regions of two operons overlap and
interfere \cite{colistat}. Among them are well-known examples such as
the \operon{lysA}-\operon{lysR} and the \operon{araBAD}-\operon{araC}
operon pairs~\cite{trpbook}.  But perhaps the best-known and arguably
the most studied example of such a motif is provided by the \lphage\
switch, which consists of two adjacent operons that mutually repress
each other \cite{lphagebook}. Here, we study a minimalist model of
such a switch, as shown in Fig.~\ref{switchfig}. In particular, we
compare the stability of an `exclusive' (XOR) switch, for which the
simultaneous binding of the repressive TFs for both operons is
inhibited, to that of a general switch. We find that the exclusive
switch is much more stable than the general switch. The basis reason
is that one cannot simultaneously turn off both operons in the
exclusive switch.  This demonstrates the potential importance of such
motifs in making gene regulatory networks robust against biochemical
noise. Although these constructions seem peculiar to prokaryotes,
evidence for correlated and anticorrelated gene expression has also
been reported for eukaryotes~\cite{WKK,Szallasi}.

The starting point of our analysis is a set of chemical reactions that
constitute the switches shown in Fig.~\ref{switchfig}.  As chemical
species, we introduce a pair of TFs which can exist as monomers,
$\TFA$ and $\TFB$, or multimers, $\TFA_n$ and $\TFB_m$.  The state of
the genome is represented by $\OO$, $\OO\TFA_n$, etc, depending on the
binding of the TF multimers.  Adopting a condensed notation in which
`$|$' indicates alternative sets of reactants and `$\hookrightarrow$'
indicates that the reactants are not destroyed by the reaction, the
set of chemical reactions are
\begin{subequations}\label{chemeqs}
\begin{align}
&n\TFA \rightleftharpoons\TFA_n,\quad
m\TFB \rightleftharpoons\TFB_m,&&(\kf, \kb)\label{eqmulti}\\ 
&\OO+\TFA_n\rightleftharpoons\OO\TFA_n,\quad
\OO+\TFB_m\rightleftharpoons\OO\TFB_m,&&(\kon, \koff)\label{eqtra}\\ 
&\OO\TFA_n+\TFB_m\;|\;\OO\TFB_n+\TFA_m
\rightleftharpoons\OO\TFA_n\TFB_m,&&(\kon, \koff)\label{eqtrd}\\
&\OO\;|\;\OO\TFA_n\hookrightarrow\TFA,\quad
\OO\;|\;\OO\TFB_m\hookrightarrow\TFB,&&(\kA),(\kB)\label{eqexpr}\\
&\TFA\;|\;\TFB\to\nothing.&&(\muA),(\muB)\label{eqdegr}
\end{align}
\end{subequations}
These reactions account for, respectively, the formation of multimers,
the binding of TF multimers to the genome (Eqs.~\eqref{eqtra} and
\eqref{eqtrd}), the expression of TF monomers, and the degradation of
TF monomers.  Repression of gene expression is implicit in
Eqs.~\eqref{eqexpr}, thus $\TFA$ is expressed if and only if $\TFB_m$
is not bound, etc.  Reaction rates are as indicated, and we define
equilibrium constants for multimerisation, $\Kd=\kf/\kb$, and binding
to the genome, $\Kb=\kon/\koff$.

Whilst detailed and biologically faithful models can be constructed as
has been done for the \lphage\ switch \cite{ARMcA, ABJS}, the above
model is intentionally `as simple as possible'.  We believe such an
approach is as important as detailed biological modelling in
elucidating the basic physics behind switches.  Thus, for example, we
have condensed the details of transcription and translation into a
single reaction step in Eqs.~\eqref{eqexpr}, governed by rate
coefficients $\kA$ and $\kB$.  On the other hand, as Cherry and Adler
have shown \cite{CA}, the TF binding isotherms must satisfy certain
criteria in order to make a working switch.  In the present model this
is effected by introducing co-operativity through the binding of TF
multimers rather than monomers.

\begin{table}
\caption{Distinct possibilities for the subsets of allowed genome
states for our switch model.\label{tab-poss}}
\begin{ruledtabular}
\begin{tabular}{lcccc}
case / genome states  & $\OO$ & $\OO\TFA_n$ & $\OO\TFB_m$ & $\OO\TFA_n\TFB_m$\\
\hline
general                & \tick  & \tick  & \tick  & \tick  \\
exclusive              & \tick  & \tick  & \tick  & \cross \\
partially co-operative & \tick  & \tick  & \cross & \tick  \\
totally co-operative   & \tick  & \cross & \cross & \tick  \\
\end{tabular}
\end{ruledtabular}
\end{table}

In our model the genome is in one of four states $\{\OO, \OO\TFA_n,
\OO\TFB_m, \OO\TFA_n\TFB_m\}$.  We now include the effect of
interference between the upstream regulatory domains by disallowing
some of these states (this is in the spirit of simplicity, strictly
speaking the effect is to modify the probabilities of the states).
Since the empty genome is always a possibility and both $\TFA_n$ and
$\TFB_m$ should be allowed to bind otherwise they would not be TFs, it
turns out that there are only five possibilities, two of which are
related by symmetry.  The four distinct cases are shown in Table
\ref{tab-poss}, and are implemented by excluding some of the reactions
in Eqs.~\eqref{eqtra} and \eqref{eqtrd}.  For example, the exclusive
switch is obtained by discarding the reactions in Eqs.~\eqref{eqtrd}
thereby removing the state $\OO\TFA_n\TFB_m$.

\begin{figure}
\begin{center}
\includegraphics{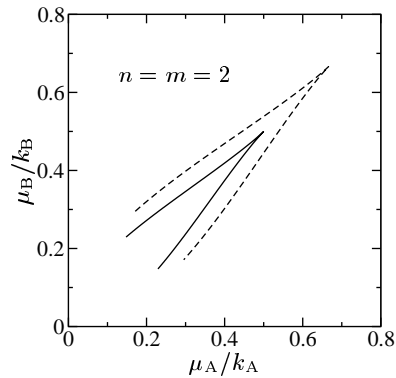}
\end{center}
\caption{In mean field theory, switching behaviour is confined to a
wedge in the $(\muA/\kA,\muB/\kB)$ plane.  Shown are the phase
diagrams for the dimerising general (solid line) and exclusive (dashed
line) switches.  It is seen that the region of bi-stability is larger
for the exclusive switch than for the general switch. For the
partially co-operative dimerising switch ($\OO\TFB_2$ disallowed), the
wedge moves to $\muA/\kA\alt0.10$ and
$\muB/\kB\alt0.019$.\label{mftheory} }
\end{figure}

We first use mean-field theory to analyse the behaviour of
Eqs.~\eqref{chemeqs}. Switching behaviour corresponds to the
appearance of two distinct stable states in the space of TF molecule
numbers.  Previously, general switches were studied in detail by
Cherry and Adler \cite{CA}, and a specific example of the exclusive
switch was studied by Kepler and Elston \cite{KE}.  We extend the
analysis of Cherry and Adler to determine where switching behaviour
can occur, for all the cases in Table~\ref{tab-poss}.  Firstly, for
$n=m=1$, no switching behaviour can be found for \emph{any} case.
This confirms that some form of co-operative binding is required.  For
the totally co-operative switch though, switching behaviour cannot be
found for \emph{any} values of $n$ and $m$.  For the remaining cases,
we have analysed in detail the situation for $n=m=2$ where both TFs
bind as dimers.  Fig.~\ref{mftheory} shows the regions in the
$(\muA/\kA, \muB/\kB)$ plane where switching behaviour is found.
Clearly, switching behaviour is more extensive for the exclusive
switch compared to the general switch, and is strongly suppressed for
the partially co-operative switch.  Thus we conclude that, at least in
mean field theory, the structure of the switch has a strong influence
on the extent of switching behaviour.

To go beyond mean field theory, we have simulated the reactions in
Eqs.~\eqref{chemeqs} using Gillespie's kinetic Monte-Carlo scheme
which generates sample trajectories appropriate to the chemical master
equation \cite{Gillespie}.  We focus on dimerising ($n=m=2$) general
and exclusive switches, and on the symmetry line $\kA=\kB=k$ and
$\muA=\muB=\mu$.  We will use the expression rate $k\approx
0.1$--$1\,\mathrm{s}^{-1}$ \cite{ABJS} as a unit of (inverse) time and
the degradation rate $\mu$ as the main control parameter.  The choice
of the rate constants is biologically motivated, in particular we
expect expression to be a slow step and the binding equilibrium to be
biased in favour of bound states~\cite{ABJS}.  For a baseline set we
use $\kf/V = \kb = \kon/V = 5\koff = 5k$ ($\Kd/V=1$ and $\Kb/V=5$),
where $V\approx2\,\mu\mathrm{m}^3$ is the cell volume \cite{ABJS}; we
assume one copy of the genome is present.

\begin{figure}
\begin{center}
\includegraphics{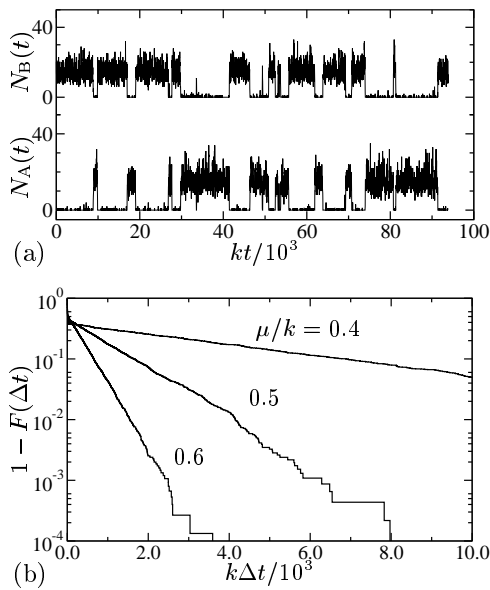}
\end{center}
\caption{(a) Typical numbers of TFs as a function of time.  (b)
Cumulative distribution functions for the time intervals between zero
crossings of $\NA-\NB$.  Results are for the dimerising exclusive
switch at $\mu/k=0.45$ unless stated otherwise.\label{signal}}
\end{figure}

We monitor the total numbers of the TFs, $\NA$ and $\NB$, including
those in dimers and those bound to the genome.  If the system is
behaving as a switch then we typically see that one of the TFs is
strongly repressed compared to the other one.  A switching event
occurs when the roles of the two TFs flip spontaneously, as shown in
Fig.~\ref{signal}.

\begin{figure}
\begin{center}
\includegraphics{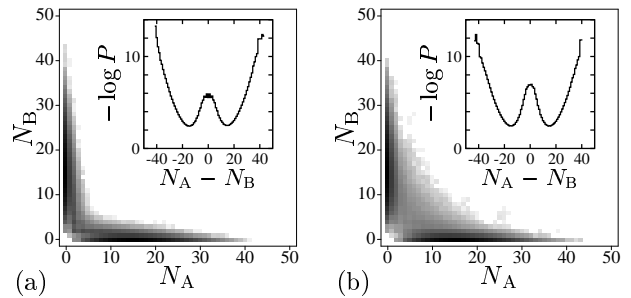}
\end{center}
\caption{Probability density in $(\NA,\NB)$ plane constructed from
$2.5\times 10^6$ samples (total duration of $kt\approx 5\times 10^6$),
for (a) general and (b) exclusive dimerising switches, at
$\mu/k=0.45$.  Greyscale indicates bin count, logarithmically, from
$\le1$ (white) to $\ge10^5$ (black).  Insets show probability density
collapsed onto the $\NA-\NB$ line, plotted as a dimensionless ``free
energy'' $-\log[P(N_A-N_B)]$ (the ordinate zero is
arbitrary).\label{densmap}}
\end{figure}

We can obtain more insight into the switching behaviour by sampling
the probability distribution $P(\NA,\NB)$ for states in the
$(\NA,\NB)$ plane, as shown in Fig.~\ref{densmap}.  Switching
behaviour appears as a double maximum in probability in this
representation, and the transition state is seen to lie at low numbers
of both TFs. Three points are worthy of note. First of all, it is seen
that the positions of the two stable steady states do not depend much
on the architecture of the switch. This is not surprising, because if
one species dominates, both switches will behave similarly. What is
perhaps more surprising, is that the pathways for switching are
different. The transition paths of the exclusive switch cross the
transition state surface at higher values of $\NA=\NB$, as compared to
the general switch. The reason is that in the general switch both
genes can be repressed simultaneously, while in the exclusive switch
only one gene can be turned off at the time. More importantly,
however, the barrier for flipping the switch is higher for the
exclusive switch than for the general switch, as can be seen in the
insets in Fig.~\ref{densmap}. This is because for a switch to flip,
two events have to happen. First of all, the system has to wait for a
rare fluctuation by which the concentration of the dominant species
decreases; this allows for the synthesis of the other
component. Subsequently, the latter component has to bind to its
operator site in order to toggle the switch. In the general switch,
the latter event is more probable, because the minor component can
bind to its site as soon as it is synthesized, while in the exclusive
switch the dominant species first has to dissociate from the DNA. This
is the main reason why the exclusive switch is more stable than the
general switch.

We have also characterised the switching {\em dynamics}, by
constructing the cumulative distribution function $F(\Delta t)$ for
the time intervals $\Delta t$ between zero crossings of the order
parameter $\NA-\NB$.  About 50\% of $F$ arises from noise on a time
scale $\Delta t \sim k^{-1}$ as the system jitters around the
transition state, but for $\Delta t\gg k^{-1}$, and provided we are
well into the switching regime, we invariably see Poisson statistics
with $F \to 1 - \exp[-\Delta t/\tau]$ (see Fig.~\ref{signal}(b)).
This firstly confirms that the switch states \emph{have} a well
defined lifetime $\tau$, and secondly allows us to extract an accurate
estimate of the value of $\tau$.

Bialek has suggested that the switch lifetime may grow exponentially
with the number of molecules involved in switching between states
\cite{Bialek}.  Motivated by this, we monitor the mean number $\nmean$
of the most-expressed TF, defined to be the time average of
$\mathrm{max}(\NA,\NB)$.  We can also calculate $\nmean$ from mean
field theory, and we find good agreement between this and the value
measured in the simulations, as $\mu$ varies.

\begin{figure}
\begin{center}
\includegraphics{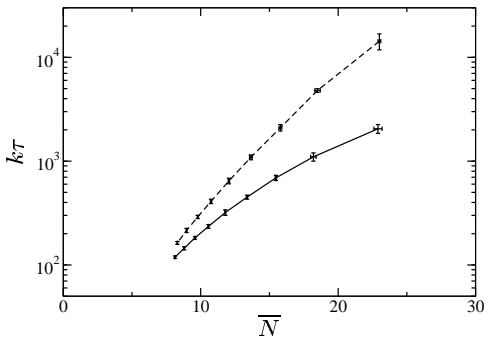}
\end{center}
\caption{Switch lifetime as a function of mean number of
most-expressed TF, for the general (solid line) and exclusive (dashed
line) cases.  The exclusive switch becomes orders of magnitude more
stable than the general switch at high numbers of the expressed
TF.\label{taunum}}
\end{figure}

Qualitative support for Bialek's conjecture comes from
Fig.~\ref{taunum}. It shows that $\tau$ grows very rapidly with
$\nmean$, which is the basic reason why extremely stable switches can
be built with at most a few hundred expressed proteins.  In contrast
to Bialek's conjecture, however, $\tau(\nmean)$ appears to be
sub-exponential in $\nmean$.  Interestingly, we can fit $\tau(\nmean)$
to a form that is suggested by an analysis of a related problem,
namely that of switching between broken-symmetry phases in a driven
diffusive model \cite{EFGM,GLEMSS}. This suggests that the ultimate
scaling is $\tau\sim \nmean{}^\alpha\exp[b\nmean]$, where $\alpha$ and
$b$ are constants. Note that this corresponds to Bialek's conjecture,
but with a logarithmic correction in $\nmean$.

More importantly, however, Fig.~\ref{taunum} clearly demonstrates that
the switch construction does indeed have a marked influence on the
stability of the switch.  It shows that the lifetime of the exclusive
switch grows much more rapidly with mean copy number than that of the
general switch~\cite{notepar}.  Our simulations cover
$10\alt\nmean\alt30$, but if we extrapolate our results to
$\nmean\approx100$, then $k\tau\approx10^4$--$10^6$ for the general
switch but $k\tau\approx10^8$--$10^{10}$ for the exclusive switch.  In
the latter case, this corresponds to lifetimes measured in tens of
years.  Such extremely long lifetimes have been reported for \lphage\
\cite{ABJS}.

In summary, a genetic switch is intrinsically stochastic, because of
the molecular character of its components. However, our simulations
demonstrate that the stability of a genetic switch can be strongly
enhanced by spatially arranging the operons such that \emph{competing}
regulatory molecules \emph{mutually exclude} each other at the
operator regions. Such a spatial arrangement can be achieved if the
two operons lie next to each other on the DNA and are transcribed in
diverging directions -- a network motif that has been identified by
our statistical analysis of the gene regulatory network of
\species{E. coli} \cite{colistat}. Hence, our simulations suggest that
robustness against biochemical noise can provide a selection pressure
that drives pairs of operons, that either regulate each other or are
controlled by a common transcription factor, towards each other in the
course of evolution.

We thank Daan Frenkel, Fred MacKintosh and Sander Tans for useful
discussions and for a critical reading of the manuscript, and Martin
Evans for drawing our attention to the work on driven diffusive
systems. This work is supported by the Amsterdam Centre for
Computational Science (ACCS). The work is part of the research program
of the ``Stichting voor Fundamenteel Onderzoek der Materie (FOM)",
which is financially supported by the ``Nederlandse organisatie voor
Wetenschappelijk Onderzoek (NWO)".


\begin{thebibliography}{19}
\bibitem[{not()}]{notepar}
\bibinfo{note}{We have also examined the effect of varying parameters away from
  the baseline set, such as increasing the number of TFs generated per
  expression reaction in order to mimic the synthesis of multiple copies of TFs
  from a single mRNA transcript. The associated changes in the noise of gene
  expression~\cite{Ozbudak} affect the switch stability, but the main
  conclusion that the exclusive switch is much more stable than the general
  switch remains unchanged.}
\end{thebibliography}

\end{document}